# Segmentation of pores in cementitious materials based on backscattered electron measurements: a new proposal of regression-based approach for threshold estimation

Andrzej M. Żak [a]*, Anna Wieczorek [b], Agnieszka Chowaniec [c], Lukasz Sadowski [c]

[a] Faculty of Chemistry, Wroclaw University of Science and Technology, Wroclaw, Poland
[b] Faculty of Pure and Applied Mathematics, Wroclaw University of Science and Technology, Wroclaw, Poland
[c] Faculty of Civil Engineering, Wroclaw University of Science and Technology, Wroclaw, Poland
* Corresponding author: andrzej.zak@pwr.edu.pl

**Abstract**

In the following work, we described the problems of porosity analysis of cement materials using backscattered electron images. We noticed that despite its great utility, the overflow porosity segmentation method allows for the introduction of an additional error by the effect of the subjective researcher. For this purpose, two developed variants of this method - regression I and II - were completely algorithmized and compared with the literature methods of overflow, triangle, and method preserving to choose the fastest and most consistent measurement method. Based on the comparison of the two data sets, it was judged that the improved overflow-based methods are the best candidates for the automated porosity assessment, with particular emphasis on regression II. All the algorithms used are summarized as Python source code in Supplementary Material.

## 1. Introduction

The methods of imaging multiphase systems reach the beginning of the popularization of scanning electron microscopy (SEM). Primary electron signals used in SEM are low-energy secondary electrons and high-energy backscattered electrons. Both are generated more efficiently for heavier atomic nuclei, but this relationship is much stronger for the second signal. The backscattering coefficient, although it does not depend perfectly linearly on the atomic number Z, allows one to easily rank the shades of gray observable in different places in the image according to the rule: the brighter the point of the image, the greater Z is associated with it [1]. This advantageous feature of the BSE image and the possibility of obtaining the 'Z contrast' very quickly interested researchers dealing with inhomogeneous and multiphase systems, including cement composites [2–5]. The method also quickly gained popularity due to the progress of automation of quantitative analysis of porosity [6] and phase composition [7,8]. Among others, SEM analysis became the basis for the recognition of individual structural compounds [9,10] of the microstructure of the cement-based material and gave the possibility of unique porosity studies within the interfacial transition zone (ITZ) between cement paste and aggregates [11]. Thanks to the further development of imaging techniques, it has become a standard to describe the macroscopic properties of a material using a microstructure and vice versa - by modelling the microstructure, we can improve the macroscopic properties [12,13].

The new method of analysis brought with it new challenges in terms of proper sample preparation. To prepare the cross section, the samples are infiltrated with resin to minimize cracking and damage to the sample during mechanical grinding and polishing [14–16]. The infiltration of open pores with resin also allows you to take advantage of the fact that the light components of the infiltrated medium have a small backscattering coefficient and together with noninfiltrated pores are responsible for the darkest parts of the image. Therefore, it becomes crucial to determine the right cut-off threshold to detect the pores. For years, a model approach has been the method described in the work [17], in which the threshold level is determined by the inflection point of the cumulative image histogram. The method has proven successful not only for the description of homogeneous samples but also for the analysis of interfaces in cement composites [18,19]. Basic works on generating the BSE signal, taking into account the depth of the beam-material interaction, are also continued [20].

Recent years have brought an extremely intensive development of improved methods of image analysis [21–24], also based on deep learning methods [25]. In addition to electron methods, light microscopy [26], as well as the intensively developed method of computed tomography [27] are other approaches used in the description of porosity. Regardless of the favorite and chosen method, the whole community has one goal - to establish an objective, repeatable, and consistent way to the quantitative description of material porosity. Our work deals with the development of the image analysis method described in the fundamental and still popular work [17]. The method requires the determination of two linear regressions of the cumulative histogram fragments. This approach proved to be useful, however, in our opinion, introduces the subjective factor of the researcher choosing a fragment of the curve. This may introduce a relatively large measurement error, which is described in detail in Chapter 2.1.1 and in Figure 2. For this reason, we decided to expand the cited method to an automated approach.

After introducing the algorithm improvements, we compared the new approach with the overflow method [17] as well as the triangle and moment preserving methods, mentioned in [18]. At the same time, we found IsoData and Otsu's methods [18] impossible to implement with a satisfactory result. The main research significance of our work is the idea of absolute algorithmization of image analysis in a way that minimizes the subjective influence of the operator on the segmentation obtained. Our approach, however, differs from that of another automatic approach [23], as we do not focus on pore segmentation and MIP matching, but on obtaining the most consistent data for the purposes of automated comparative analysis of large volumes of data. For easy access to the mentioned algorithms, they are all summarized in the Supplementary Material for easy use.

## 2. Materials and methods

The samples were made from CEM III/A 42,5N-LH/HSR/NA cement (300 kg/m$^3$), fly ash (55 kg/m$^3$), and two different water-cement ratios (w/c) – 0.60 (sample I) and 0.48 (sample II). The addition of fly ash provides greater homogeneity of the mix, which facilitates quantitative analysis of porosity for statistical comparison of methods. The cement paste was cured for 30 days and cut into 1 cm$^3$ blocks to minimize the influence of the sample on high-vacuum devices. The samples were dried at 50 ° C for 72 h until weight loss stopped, vacuum embedded in low viscosity Epidian 652 epoxy resin (Ciech Sarzyna, Poland), ground to 1200 grit, and mechanically polished with 6 μm and 1μm diamond paste to the surface roughness of the surface roughness level of $S_a$=300 nm. To obtain the electric conductivity of the specimen and avoid electric charging of the specimen, a copper tape-made conductive track is applied and the whole specimen is covered with a 40 nm thick layer of graphite coating in a high-vacuum thermal evaporator. SEM imaging was performed using a JEOL JSM-6610A scanning electron microscope (Tokyo, Japan), using the two-field BSE detector, an accelerating voltage of 20 kV, gun current of 40 nA, and a working distance of 10 mm. Micrographs with a resolution of 1280x960 pix were taken with the longest available acquisition time, 180 s. For each sample, 30 photos were taken at a magnification of 200x, which allowed for a resolution of 0.5 μm/pixel and a total imaging coverage area of 8.4 mm$^2$ for each sample. The brightness and contrast settings have been adjusted so that the histogram covers the entire grayscale, but without peaks in extreme black and white (0, 255). Each of the images was subjected to separate thresholding to make the result independent of the image order. With the Kolmogorov-Smirnov test for each data series, we tested that the distribution of values corresponds to a normal distribution with a confidence level of at least 95%.

Below we compile and describe mathematically three existing methods - triangle [28], moment preserving [29], and overflow [17], which, after our modification, created two additional variants, called regression I and II. They were compared in the experimental part of the work.

We start with an image taken with backscattered electrons using material contrast in a scanning electron microscope represented by an array $M = [m_{ij}] \in \mathcal{M}_{I \times J}$, where $m_{ij} \in [0,255]$ are pixels and I and J are image dimensions in pixels. We create a frequency histogram of the occurrence of each grey value in the image according to the formula $h(k) = \sum I \sum J \mathbb{1}_{\{m_{ij}=k\}}, k \in [0,255]$, and a cumulative histogram $H(k) = \sum I \sum J \mathbb{1}_{\{m_{ij} \leq k\}}$. To simplify notation, we use the notation $H_k$, which is equivalent to $H(k)$. For a comparative presentation, all the methods below have been explained with the example of the photo in Figure 1.

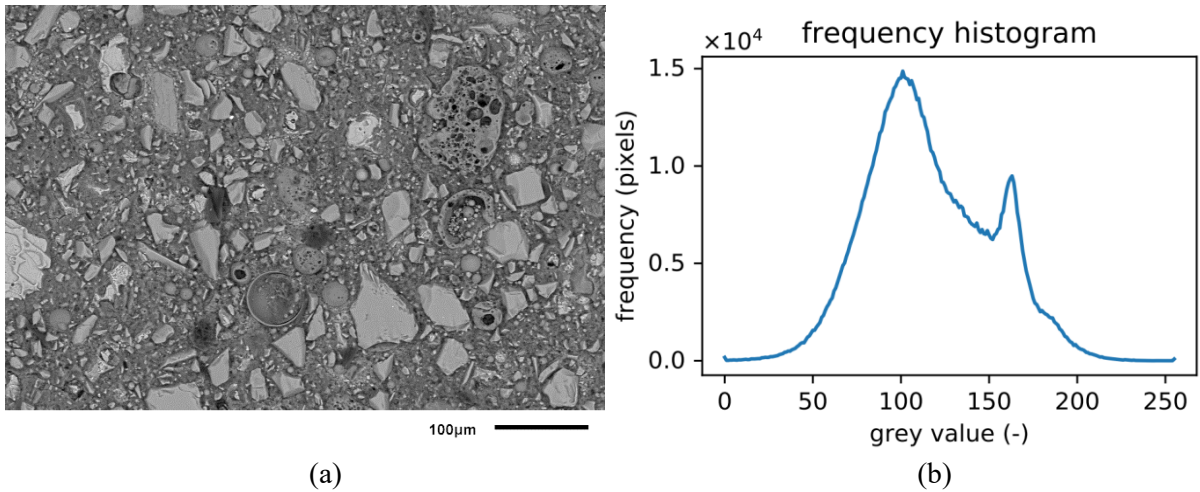

**Figure 1.** BSE image (a) and its brightness histogram (b).

### 2.1. Existing threshold determination methods

#### 2.1.1. Overflow pore segmentation method [17]

The method was developed based on the observation that the proper threshold level for the porosity could be represented by the point where a further grey value growth will cause a sudden increase in the thresholded area. As defined in the article [17], the threshold reflects to the inflection of the cumulative histogram curve, which can be estimated from the intersection of two linear sections of the cumulative histogram. Figure 5 of article [17], in particular, the fit coefficient of the regression model $r^2$, for the construction of the linear sections suggests the use of a linear regression model. Unfortunately, the method does not specify which sections of the histogram to use, or how to determine that a section is sufficiently 'linear'. As a result of this misspecification, within the same cumulative histogram, a researcher considering different subjective measures of linearity may construct many different linear sections that may determine different thresholds (Figure 2), varying in this example from 66 to 70. The threshold value of 69 (Figure 2cg) provides the lowest $R^2$ values of the examples mentioned, however, when using different parts of the curve, these values may change, resulting in a difference in the threshold value. Even such a small difference in the threshold value, within the ±2 value limit, causes the boundary values of the segmented porosity to range from 5.85 % to 7.72 %, which in this case gives a relative value change of over 30 %. This means that the researcher may unintentionally introduce an error greater than the difference in the compared images or samples.

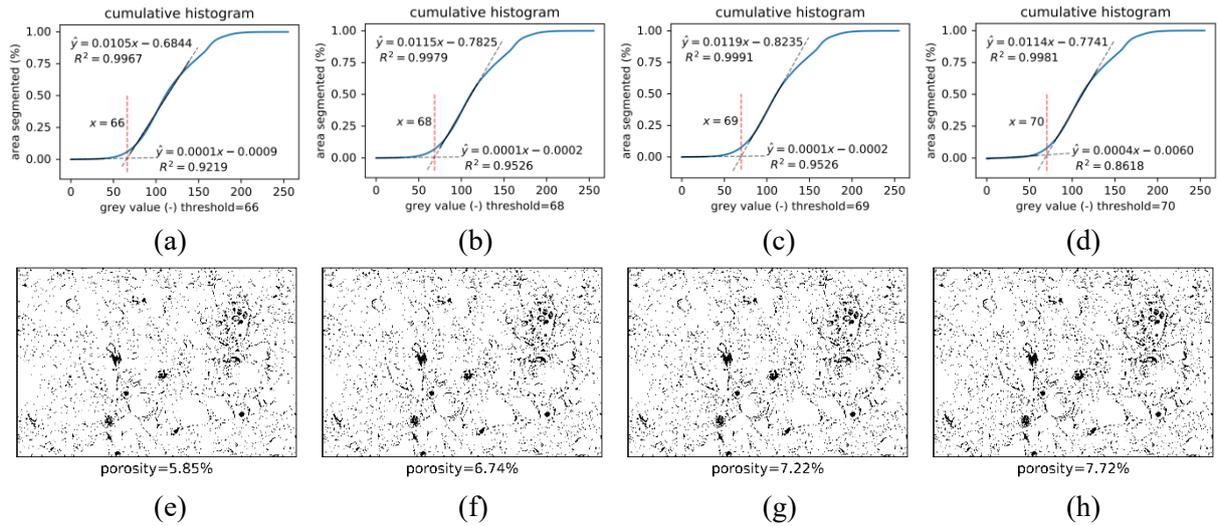

**Figure 2.** Manual overflow method: different thresholds, for different straight sections of cumulative histogram (a-d) and representative pore segmentation (e-h)

### 2.1.2. Modified triangle method [28]

To determine the threshold, we can also use the triangle method [28] modified for our problem. We perform the fit on a cumulative histogram bounded as in the regression I method to arguments from $a$ (the origin of the cumulative histogram) to $b$ (the argument corresponding to the peak of the hydration products). The method consists in finding the cumulative histogram point $(k, H_k)$ that is the furthest from the straight line passes through points $(a, H_a)$ and $(b, H_b)$.

Thus the threshold $t^*$ maximizes the function $d(k)$ (equation 1), it is noteworthy that it is sufficient to look for the maximum of the numerator because the denominator is always a positive constant. A graphical representation of the method is shown in Figure 3.

$$d(k) = \frac{|(H_b - H_a)(k-a) - (b-a)(H_k - H_a)|}{\sqrt{(b-a)^2 + (H_b - H_a)^2}}. \tag{1}$$

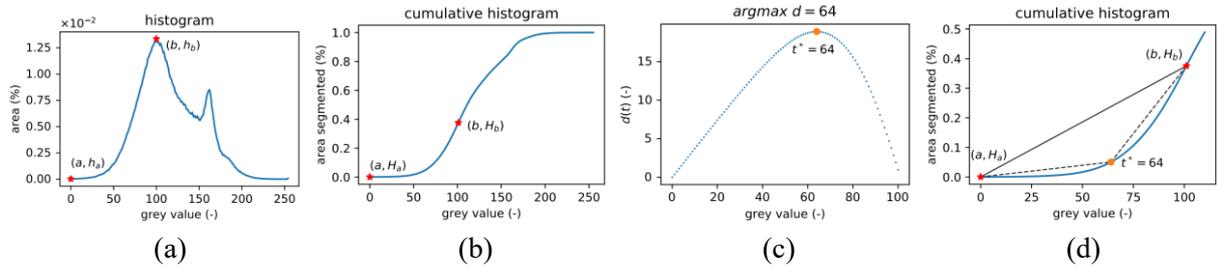

**Figure 3.** Modified triangle method: (a) histogram; (b) cumulative histogram; (c) function d(k); (d) part of the cumulative histogram with the longest height triangle marked. In the graphs, red stars mark the start and end points, and an orange dot marks the $t^*$ threshold.

### 2.1.3. Moment preserving method [29]

We also used the moment preserving method very well described in the paper [29]. As with the previous methods, in order for the method to separate the pores well, we need to limit the portion of the histogram we are using. Since the point we are looking for is between zero and the peak of the hydration products. (like the red stars in Figure 1), we will use these points as limits.

### 2.2. Development of a regression threshold method for assessing porosity in cementitious materials

### 2.2.1. Minimizing the error of fitting straight lines to a graph (regression I)

The method for minimizing the error of fitting straight lines to a graph is based on a linear least squares regression model. When given a set of observed data $\{x_i, y_i\}_{i=1}^n$ fitting a straight line $\hat{y} = ax + b$ reduces to the formula:

$$\begin{cases} a = \frac{\sum_{i=1}^n y_i(x_i - \bar{x})}{\sum_{i=1}^n (x_i - \bar{x})^2}, \\ b = \bar{y} - a\bar{x}, \end{cases} \quad (2)$$

where $\bar{x} = \frac{1}{n}\sum_{i=1}^n x_i$, $\bar{y} = \frac{1}{n}\sum_{i=1}^n y_i$ are sample averages. We express the fitting error by the errors sum of squares $SSE = \sum_{i=1}^n (y_i - \hat{y}_i)^2$.

Modifying the above method, we can use it to determine a threshold (consistent with the overflow method [17]. First, we need to determine the part of the cumulative histogram to which we will fit the two regression lines. Because of the simplicity of the determination, the associated repeatability of the measurement, and the possibility of some automation of the process, we take the beginning of the cumulative histogram as the starting point; let us denote this point as $a$. The final point -- the inflection point of the cumulative histogram, understood as the point where the function changes from being concave to convex, which is at the same time the global maximum of the histogram (the peak of the hydration products), let us denote this point as $b$.

To find the threshold $t^*$, we determine the error statistic that we want to minimize by fitting straight lines to the cumulative histogram, with intent similar to work [17], but with a different approach. In our case it will be the sum of the fitting errors of the first and second curves, we denote it $E(t)$. For each $t \in (a, b)$ dividing the piece of the cumulative histogram of interest $\{k, H_k\}_{k=a}^b$ into two sets $\{k, H_k\}_{k=a}^{t-1}$ and $\{k, H_k\}_{k=t}^b$, we fit regression lines $\widehat{y_1}$ and $\widehat{y_2}$, respectively, and compute the errors $SSE_1$ and $SSE_2$ obtaining the error at $t$:

$$E(t) = SST_1 + SST_2 = \sum_{i=a}^{t-1}(H_i - \widehat{y_{1i}})^2 + \sum_{i=t}^{b}(H_i - \widehat{y_{2i}})^2. \quad (3)$$

The threshold $t^*$ is the argument of the function $E$ where it reaches its minimum. A graphical representation of the method is shown in Figure 4.

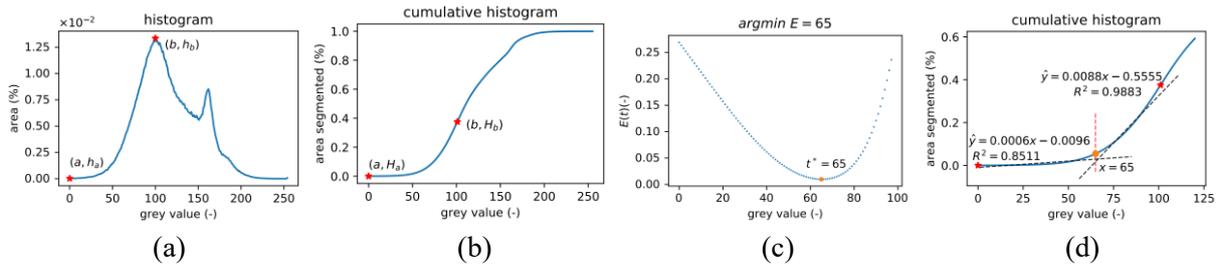

**Figure 4.** Regression I method: (a) histogram; (b) cumulative histogram; (c) E(t) function; (d) part of the cumulative histogram with fitted regression lines marked. In the graphs, red stars indicate the start and end points, and an orange dot indicates the $t^*$ threshold.

### 2.2.2. Fit around an inflection point (regression II)

We can also use linear regression in another way. We take point b corresponding to the argument of the peak of the hydration products, and extract two sets from the cumulative histogram: $\{k, H_k\}_{k=0}^{b/4}$ and $\{k, H_k\}_{k=5b/4}^{5b/4}$, to which we fit straight lines using equation 2. Then we can calculate the intersection point of the lines obtaining the threshold $t^*$. A graphical representation of the method is shown in Figure 5.

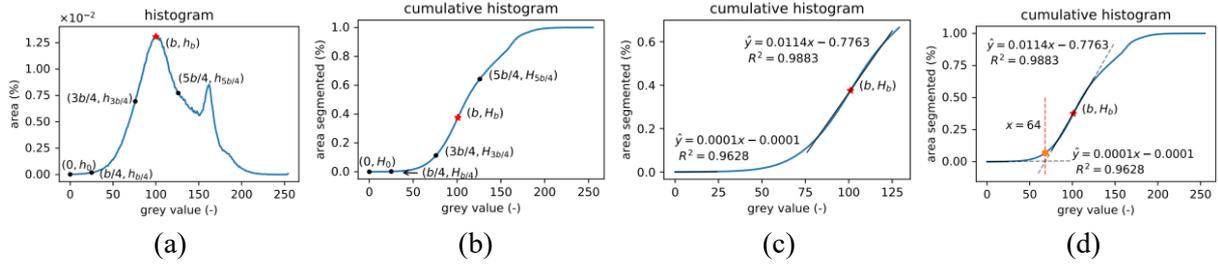

**Figure 5.** Regression II method: (a) histogram; (b) cumulative histogram; (c) part of the cumulative histogram with fitted regression lines marked; (d) cumulative histogram with the orange dot indicating the $t^*$ threshold. In the graphs, red stars indicate the b point, and the black dot marks the start and end points of the sets.

## 3. Experimental results

The problem of comparing the methods of porosity segmentation is a demanding task. The true value of the porosity remains unknown, however, by dividing the sample into smaller fragments, we obtain local results that cumulatively reflect the entire area studied. Note that porosity analysis based on SEM-BSE images has some limitations. We cannot see pores smaller than the pixel size (in our case 0.5 µm), and the macro-voids occupying most of the frame will be a significant disturbance of the algorithm. For this reason, data acquisition must still be monitored. In addition to the visual control of the convergence of the cumulative porosity score, we tried to introduce the additional measures of the algorithm comparison – $N$ and $S^2$.

For both samples, marked I and II, a cumulative porosity analysis was performed based on a group of 30 images, and the statistical results are summarized in Table 1. Since the precision is half the length of the 95 % confidence interval, $N$ is the number of samples needed to obtain a measurement with a precision of 0.1, calculated from the formula

$$N = \left\lceil S^2 \left(\frac{t_{0.975,29}}{0.1}\right)^2 \right\rceil, \tag{4}$$

where $t_{0.975}$ is percent point function (inverse of CDF) of t-student distribution with 29 degrees of freedom in point 0.975 (equal to 2.045); 0.1 is the precision we require; and $S^2$ is variance.

In both cases, the highest porosity was read using moments preserving methods, and the smallest for the triangle method. Developed methods and the overflow method were in between, but it should be borne in mind that none of the methods mentioned above gives an objective measurement of porosity. However, we want the measurement method to be as objective as possible and able to work with the smallest possible data sets. For this, it is worth paying attention to the N needed to obtain a measurement with a precision of 0.1. The preserving method requires the largest datasets, 45 and 51. The overflow method required 35 and 25 images, and triangle 28 and 13, so the least of the literature methods. In both cases, the developed method required the smallest data sets, for regression I – 23 and 9 microstructures respectively, and 16 and 7 images for regression II. Although in samples I and II the N values differ even several times, the methods of analysis are arranged in the same order. This means that the proposed methods are a significant improvement of the overflow method, with an additional indication of the regression II method.

Equally important features are indicated by the difference of the variation obtained with the above-mentioned methods on the same sets of 30 images. For sample I, the variance ranges from 0.365 % (regression II) through 0.548 % (regression I) up to 1.064 % (moment preserving). For sample II, the differences are much more vivid, and the variance ranges from 0.097 % (regression II), through 0.203 % (regression I) up to 1.210 % (moment preserving). The triangle and overflow methods are in between regression I and moment preserving, but the lowest variance of the proposed methods means that the researcher is more confident when comparing different samples or data sets. Of course, if it is necessary to compare the results of porosity segmentation with other methods, such as mercury intrusion porosimetry, it is worth using the method described in [23]. However, when performing a comparative analysis of a set of samples, we aim to minimize the impact of the method on the obtained numerical value, and in this case, we recommend using regression II or I methods or preferably comparing all the attached methods on the data set of interest to us.

**Table 1.** Comparison of porosity segmentation by different methods, average $\mu$, variance $S^2$, confidence intervals $CI$, minimum $min$, maximum $max$ and number of samples $N$ needed to obtain a measurement with the precision of 0.1.

|  | μ (%) | CI (%) | S² (%) | min (%) | max (%) | N (-) |
|---|---|---|---|---|---|---|
| **Sample I,** w/c=0.60 | | | | | | |
| overflow | 6.961 | [6.653, 7.270] | 0.827 | 5.180 | 9.648 | 35 |
| triangle | 5.304 | [5.056, 5.552] | 0.665 | 3.795 | 7.802 | 28 |
| moment preserving | 8.799 | [8.402, 9.196] | 1.064 | 6.297 | 10.742 | 45 |
| regression I | 5.814 | [5.610, 6.019] | 0.548 | 4.472 | 8.235 | 23 |
| regression II | 6.804 | [6.668, 6.941] | 0.365 | 5.801 | 8.684 | 16 |
| **Sample II,** w/c=0.48 | | | | | | |
| overflow | 6.562 | [6.347, 6.776] | 0.574 | 5.115 | 7.860 | 25 |
| triangle | 4.741 | [4.633, 4.850] | 0.291 | 3.970 | 5.901 | 13 |
| moment preserving | 9.355 | [8.903, 9.807] | 1.210 | 6.819 | 11.661 | 51 |
| regression I | 5.421 | [5.345, 5.497] | 0.203 | 4.707 | 6.418 | 9 |
| regression II | 6.419 | [6.383, 6.455] | 0.097 | 5.967 | 7.091 | 5 |

The function of the cumulative porosity measurement from the randomly arranged data shows that the measured porosity values for the regression I, II, and triangle methods stabilize around 10-15 images (Fig. 6ab). The overflow method for sample II behaves similarly, but for sample I, the read value decreases successively throughout the data set. This is related to images 2-4 which significantly increased the average porosity (Fig. 6a). However, it should be noted that some of the methods dealt with this disturbance much faster. The initial course of data from sample II behaves similarly, which is typical for multiphase cement composites and relatively unrepresentative magnifications of observations. The comparison of the cumulative variance courses (Fig. 6cd) confirms that even for very small and unrepresentative data volumes, the regression methods show the lowest variability. The highest variances are typical for moment preserving and overflow. In the latter case, the likely reason is slight subjectivity in selecting the appropriate curve sections for linear regression, described before in Section 2.1.1.

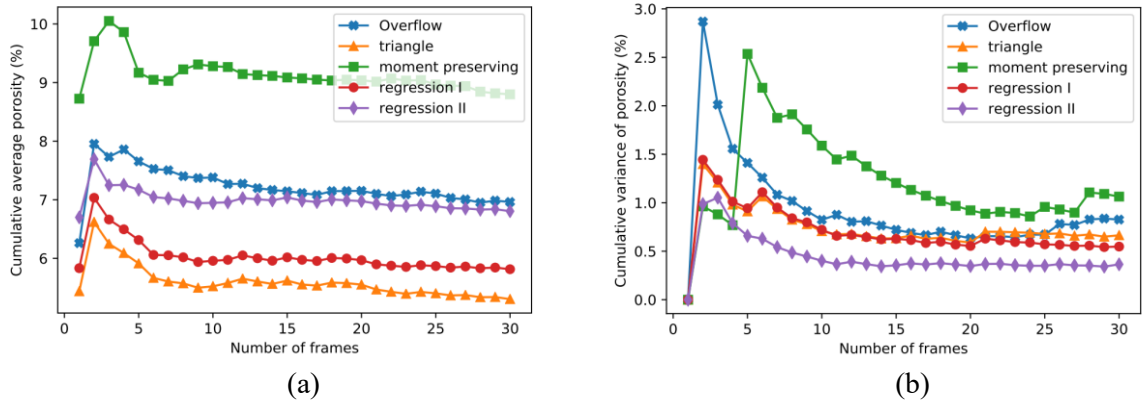

(a)      (b)

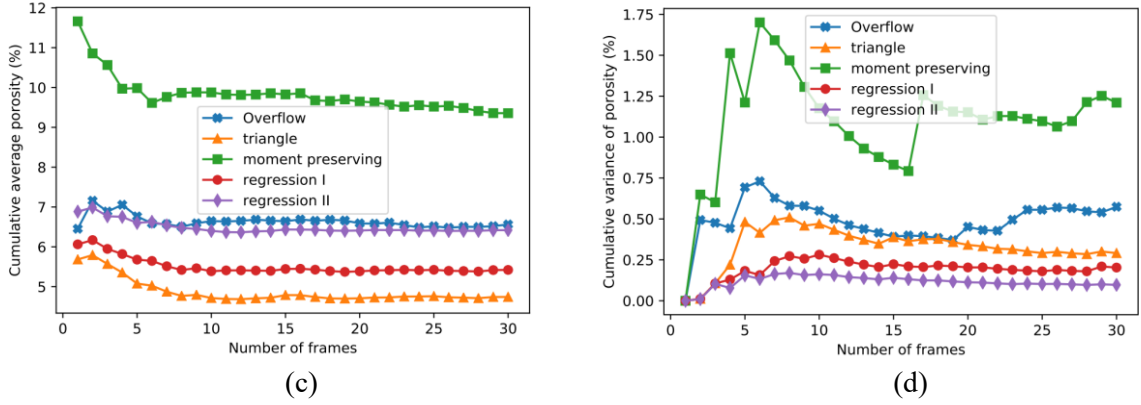

Figure 6. Cumulative average porosity (a,b), and cumulative variance porosity (c,d).
Sample I, w/c=0.60 (a,c) and Sample II, w/c=0.48 (b,d).

## 4. Conclusions

In the above work, we achieved our aim to develop operator-independent algorithms for the assessment of porosity based on SEM-BSE images, and then we compared their performance on two data sets. We have shown that both literature and development methods show a significant difference in the number of samples N required to achieve the same consistency and variance on the same data set. All the algorithms used have been made available for use in the Python environment in Supplementary Material. Based on the results obtained, the following conclusions can be made:

1. The smallest variances and the smallest required volume of data were achieved using the developmental methods regression II and I, then using the literature methods triangle [28], overflow [17], and the least consistent moment preserving methods [29].
2. The above analysis does not have to reflect the actual porosity of the sample and depends on the selected number of photos and their magnification, however, it is based on the well-established overflow method [17], further improving its consistency.
3. Automatic and algorithmized methods of regression II and I, despite being based on the same theoretical foundations as the overflow method [17], allow to reduce the amount of data or increase the consistency of the analysis, excluding human error when selecting the threshold for pore segmentation. This can help quantify discipline-relevant issues, such as refinement of the porosity with reaction time and its dependence on various additives and hydration conditions.
4. In the case of the possibility of consistent and repeatable detection of porosity, the next step may be to use a similar approach for the automatic detection of aggregates or another structural elements in the ITZ [30] on the microstructures of mortars and concretes. This should be the following subject of further methodology research.

**Funding**

Łukasz Sadowski received funding from the project supported by the National Centre for Research and Development, Poland — Grant No. LIDER/35/0130/L-11/19/NCBR/2020 "The use of granite powder waste for the production of selected construction products."

# Segmentation of pores in cementitious materials based on backscattered electron measurements: a new proposal of regression-based approach for threshold estimation

Andrzej M. Żak [a]*, Anna Wieczorek [b], Agnieszka Chowaniec [c], Lukasz Sadowski [c]

[a] Faculty of Chemistry, Wroclaw University of Science and Technology, Wroclaw, Poland
[b] Faculty of Pure and Applied Mathematics, Wroclaw University of Science and Technology, Wroclaw, Poland
[c] Faculty of Civil Engineering, Wroclaw University of Science and Technology, Wroclaw, Poland
* Corresponding author: andrzej.zak@pwr.edu.pl

## Supplementary material

Below we present the algorithms used in the article. The code was written in Python 3.7.3. In the first part of code we import the required libraries, initialize class AssesingPorosity and implement method needed in initialize, allow change *a* and *b* parameters and wrapper method so that the threshold calculation methods also determine the porosity.

```python
1  from PIL import Image
2  import numpy as np
3  import matplotlib.pyplot as plt
4
5  class AssesingPorosity:
6      def __init__(self,path):
7          self.array = self.__image2array(path)
8          self.hist = self.__hist()
9          self.Hist = np.cumsum(self.hist)
10         self.a = np.nonzero(self.hist)[0][0]
11         self.b = np.argmax(self.hist)
12
13     def __image2array(self,path,cut_frame=True):
14         """Turn the image into a two-dimensional numpy array. Our images
15            had a 90 pixel frame at the bottom, if yours is different you
16            can trim it manually and set 'cut_frame' to False"""
17         with Image.open(path) as image:
18             im_arr = np.frombuffer(image.tobytes(), dtype=np.uint8)
19             if cut_frame:
20                 return im_arr.reshape((image.size[1], image.size[0]))[:-90,:]
21             else:
22                 return im_arr.reshape((image.size[1], image.size[0]))
23
24     def __hist(self):
25         """Return histogram"""
26         return np.histogram(self.array, bins=np.arange(257))[0]
27
28     def set_a(self,a):
29         """Set a (start point) if it's needed."""
30         self.a = a
31
32     def set_b(self,b):
33         """Set b (end point) if HP-peak isn't global maximum."""
34         self.b = b
```

```
35
36      def porosity(method):
37          """Wrapper for treshold method, which causes them
38                  to return tuple (treshold,porosity)."""
39          def wraper(self):
40              t = method(self)
41              p = 100*self.Hist[t]/self.Hist[-1]
42              return t, p
43          return wraper
```

### 1. Modified triangle method

```
1       @porosity
2       def triangle(self):
3       """Modifiied triangle method described at section 2.1.2 of main article."""
4           k, Hk = np.arange(self.b), self.Hist[:self.b],
6           Ha, Hb = self.Hist[self.a], self.Hist[self.b]
7       return np.argmax(np.abs((Hb-Ha)*(k-self.a)-(self.b-self.a)*(Hk-Ha)))+self.a
```

### 2. Moment preserving method

```
1     @porosity
2     def moment_preserving(self):
3     """Moment preserving method described at section 2.1.3 of main article."""
4         p = self.hist[self.a:self.b]
5         p = p/sum(p)
6         z = np.arange(self.b-self.a)
7         mi = lambda i: sum(p*z**i)
8         m0, m1, m2, m3 = 1, mi(1), mi(2), mi(3)
9
10        cd = m0*m2 - m1**2
11        c0 = (m1*m3 - m2**2)/cd
12        c1 = (m1*m2 - m0*m3)/cd
13
14        z0 = 0.5*(-c1 - np.sqrt(c1**2-4*c0))
15        z1 = 0.5*(-c1 + np.sqrt(c1**2-4*c0))
16
17        pd = z1-z0
18        p0 = (z1-m1)/pd
19        p1 = 1-p0
20
21        return np.argmin(np.abs(p0-np.cumsum(p))) + self.a
```

## 3. Minimizing the error of fitting straight lines to a graph (regression I)

Note, that support function *__regression* is used also in regression II method in the next section.

```python
1    def __regression(self,x,y):
2        """Support function, return tuple with fitted line x, y
3           and a, b coefficients in order (x,y,a,b). """
4        m = np.mean(x)
5        a = np.sum(y*(x-m))/np.sum((x-m)**2)
6        b = np.mean(y) - a*m
7           x, a*x+b, a, b
8
9    @porosity
10   def regression_I(self):
11       """Regression I method described at section 2.2.1 of main article."""
12       Hist = self.Hist[self.a:self.b]
13       n = self.b-self.a
14       E = -1*np.ones(n-3)
15       for k in np.arange(2,n-1):
16           x1,x2 = np.arange(1,k+1),np.arange(k+1,n+1)
17           y1,y2 = Hist[:k],Hist[k:]
18           SSE1 = np.sum((y1 - self.__regression(x1,y1)[1])**2)
19           SSE2 = np.sum((y2 - self.__regression(x2,y2)[1])**2)
20           E[k-2] = SSE1+SSE2
21       return np.argmin(E) + self.a
```

## 4. Fit around an inflection point (regression II)

```python
1    @porosity
2    def regression_II(self):
3        """Regression II method described at section 2.2.2 of main article."""
4        n = int((self.b-self.a)/4)
5        x1, x2 = np.arange(self.a,n), np.arange(self.b-n, self.b+n)
6        y1, y2 = self.Hist[self.a:n], self.Hist[self.b-n : self.b+n]
7        x,y, a1,b1 = self.__regression(x1,y1)
8        x,y, a2,b2 = self.__regression(x2,y2)
9        return int((b2-b1)/(a1-a2)) + self.a
```

To check if HP peak is global maximum (if not it is needed to manually change b parameter) we used the following function which plot histogram, a and b parameter.

```python
1    def plot_check_graph(self):
2        """Show frequency histogram with marked a and b points.
3           Useful to check if HP peak is global maximum"""
4        plt.plot(np.arange(256),self.hist)
5        plt.plot(self.a, self.hist[self.a], '.', color='red', markersize=15)
6        plt.plot(self.b, self.hist[self.b],'.', color='red', markersize=15)
7        plt.title("frequency histogram")
8        plt.xlabel("grey value (-)")
9        plt.ylabel("frequency (pixels)")
10       plt.show()
```